# Knowing Your Population: Privacy-Sensitive Mining of Massive Data

Pedro Sanches[1], Eric-Oluf Svee[2], Markus Bylund[3], Benjamin Hirsch[4] & Magnus Boman[1]

[1] Kungliga Tekniska Högskolan (KTH), The School of Computer Science and Communication (SCS), Stockholm, Sweden

[2] Stockholm University, Department of Computer and Systems Sciences (DSV), Stockholm, Sweden

[3] Swedish Institute of Computer Science (SICS), Stockholm, Sweden

[4] EBTIC / Khalifa University Abu Dhabi, United Arab Emirates

Correspondence: Pedro Sanches, Kungliga Tekniska Högskolan (KTH), The School of Computer Science and Communication (SCS), Stockholm, Sweden. E-mail: sanches@sics.se

**Abstract**

Location and mobility patterns of individuals are important to environmental planning, societal resilience, public health, and a host of commercial applications. Mining telecommunication traffic and transactions data for such purposes is controversial, in particular raising issues of privacy. However, our hypothesis is that privacy-sensitive uses are possible and often beneficial enough to warrant considerable research and development efforts. Our work contends that peoples' behavior can yield patterns of both significant commercial, and research, value. For such purposes, methods and algorithms for mining telecommunication data to extract commonly used routes and locations, articulated through time-geographical constructs, are described in a case study within the area of transportation planning and analysis. From the outset, these were designed to balance the privacy of subscribers and the added value of mobility patterns derived from their mobile communication traffic and transactions data. Our work directly contrasts the current, commonly held notion that value can only be added to services by directly monitoring the behavior of individuals, such as in current attempts at location-based services. We position our work within relevant legal frameworks for privacy and data protection, and show that our methods comply with such requirements and also follow best-practices.

## 1. Introduction

*1.1 Problem*

Human mobility patterns may be plotted geographically or otherwise visualized to quickly grasp where people are, when they are there, and for what purpose. In the future massive data sets will be employed to improve the quality of such visualizations, serving governments, private companies, and individuals seeking to plan for, and to understand, human movement (Baraniuk, 2011). The problem to address for any such endeavor—either for commercial or societal benefit - is how to achieve the goal of *knowing your population*.

*1.2 Significance*

We make use of data generated within mobile communication infrastructures, data which also includes geospatial and demographic information about subscribers. During the last 15 years there has been increasing interest from governmental and private organizations, in many places around the world, in using such data outside its normal network uses for surveillance purposes (Becker et al., 2011; European Parliament, 2006; Giannotti & Pedreschi, 2008; Lyon, 2007, p. 43). Simultaneously, others have detected, and in some cases also realized, the marketing and research potential of such data. This shift in considerations, from operational purposes (enabling mobile communications) to surveillance and data mining has generated discussion under privacy and data protection discourses ('Data Protection Working Party, Article 29', 2009).

*1.3 Hypothesis*

The shift from operational purposes to surveillance and data mining is divisive and controversial. The possible added value and the recent advances in the refinement of massive data make the shift impossible to ignore, however, and our hypothesis is that privacy-sensitive exploitation is possible and often beneficial enough to warrant considerable research and development efforts. Our work contends that peoples' behavior can yield patterns of significant commercial and research value when viewed through the lens of time geography as aggregations. Representing mass behavior (macro-level aggregation) through time geography constructs opposes the current, commonly held notion that value can only be added to services by representing individual behavior (micro-level aggregation).

**2. Background**

*2.1 Network Data*

Mobile networks are heterogeneous assemblages composed of technical, societal, legal, political, and economic entities that first create and then hold the network together. The operator of such a network is a legal entity comprised of units such as network infrastructure, billing, services, and content provisioning. It is impossible to dissociate the technical components and systems that form such a network from the institutional and legal structures required for its ongoing development, maintenance, and use. Data flows corresponding to information about subscribers, billing, and locations of the subscribers are the structures that typically pertain to mobile operator business units. Such circulating data is what we refer to as *network data*: it is typically generated passively within the mobile network infrastructure and it is what allows calls and text messages to reach their destination and be billed over any distance. At the moment, there is a wealth of such data stored behind the firewalls of each operator, most of it linkable to a particular subscriber.

As its owners, mobile operators have been among the first to assess the potential of the data they house. For example, Docomo announced a project (Tech in Asia, 2011) where they would use aggregated mobility data from their cell phone customers in Tokyo for purposes of urban planning and earthquake preparation; BT has used call networks in the UK for similar purposes (Eagle, Macy, & Claxton, 2010); SkyHook (Webb, 2011) started their own mobility data mining projects, relying on their own sets of anonymized data, for research and marketing purposes.

The influence and interconnection that mobile telecommunication networks assert, with their socio-technical nature being enmeshed deeply into many aspects of everyday life, calls for a design process that acknowledges and incorporates organizational, economic, and societal aspects as much as the technical features (Kiron, Ferguson, & Prentice, 2013). We argue that during the current transitional stage in their development, the design of systems that make use of network data for human mobility research should be exploratory and take into account non-functional aspects, like the association of the technology in question with ethical values or appropriateness of the technology for a certain social context, as much as functional ones. Because we are at the brink of a shift in how communication network data is handled (Crovitz, 2013), the quality of the first systems making use of this kind of data will likely drive public acceptance of future systems and influence restrictions and legislation on how this data is to be used: the recent case of Apple secretly storing a positioning log file in the device of each user is likely to influence future laws dealing with such data (Bradshaw & Palmer, 2011). At the moment, there is an open design space for systems of this kind and we, as designers of new technology, have the opportunity to engage stakeholders in the co-creation of new systems and services, while paying close attention to ethical and social implications.

*2.2 Positioning*

The core network of a Global System for Mobile Communications (GSM) is comprised of a number of base stations which define cells, each uniquely identified by a cell-id, with a certain radius and geometry. Every base station is connected to a base station controller, which manages all the radio resources of its controlled base stations. To optimize routing, several base station controllers are grouped into Location Areas covering larger geographical regions (Rahnema, 1993). When the mobile device is in an active call, transmitting or receiving data, the base station controller defines which cell the device is connected to. If the controller rules that a device should switch to another cell—which can happen for a number of reasons: the subscriber may have moved, the signal strength may have faded, management of cell load factor, etc.—it initiates a handover procedure connecting the device to a new cell. The area covered by a cell varies depending on the capacity, demand, and the geographical topology where the cell is located. In an urban area, the size of one cell is generally between 100 meters and up to a few kilometers, and in rural and less inhabited areas the radius can be up to 35 kilometers. The accuracy is thus dependent on the topology of the network, for any particular subscriber.

The mobile network is able to roughly track where the subscribers are located geographically to be able to route calls and data traffic to and from them. Deemed *passive positioning*, this requires neither software to be installed on the mobile devices, nor extra network traffic to be generated: it is simply part of normal system operations. When the mobile device is idle, it continuously monitors the signal strength of nearby cells, performing a cell reselection when necessary. This procedure is usually only registered on the device. However, if the device crosses a Location Area region, it must update the network as dictated by protocol, creating a record of the base station the device is connected to. In addition, even if the device does not change Location Area, the mobile network pages the device at a set time interval that depends on the network (e.g., once per hour, cf. Milinski, 2012), creating a similar record. In short, the data generated in the network is characterized by:

1) Sequences of cell-ids corresponding to active calls, text messages, or data traffic;
2) Cell-ids corresponding to Location Area changes when the device is idle;
3) Cell-ids generated from network paging, done at regular times.

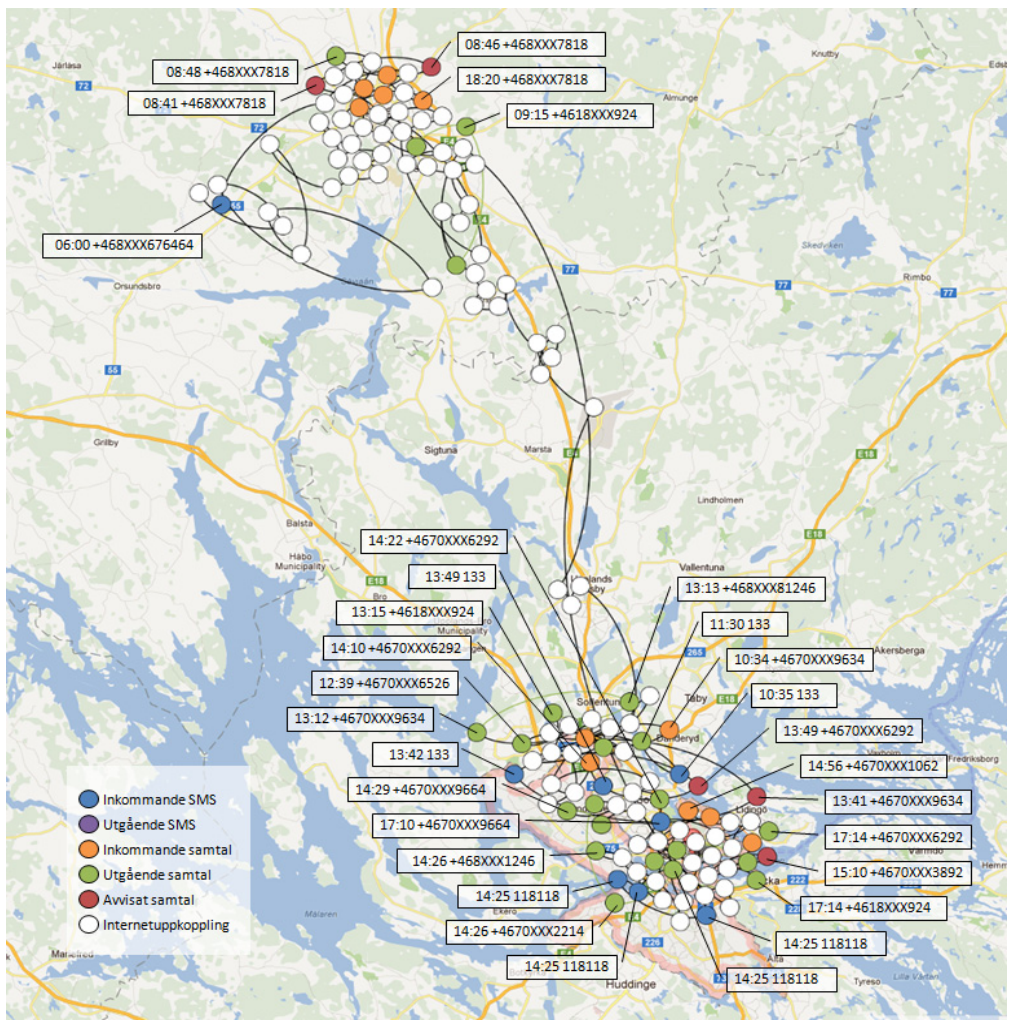

Figure 1. Annotated movement patterns of an individual subscriber, superimposed on a map of Stockholm. Colors indicate, respectively: text (incoming *blue*, outgoing *purple*), calls (incoming *orange*, outgoing *green*, rejected *red*), and internet connections (*white*)

Network data is generated in several systems across the network and, while implementations vary, is generally accessible from a system called the home location register. It is normally only stored for the period of time necessary for the purposes of handling the communication and billing the subscriber, and until recently, the only exception to this was in the event of lawful interception, where law enforcement entities may request data about a subscriber to be retained for analysis or as evidence. Data retention laws, however, require data about all subscribers to be retained for set periods of time, in many countries. Besides this data, the subscriber databases residing at operators typically hold information about each subscriber which is tied to the contract between the subscriber and the operator. This demographic data varies but generally includes name, age, and contact information, such as home address.

Creating mobility patterns from passively generated network data requires mapping each cell-id in the network to geographical coordinates. There are, however, several issues that need to be taken into account: (1) accuracy of the positioning may be low in some cases; (2) the frequency of cell-id updates depends on the activity of the subscriber and is often low, producing sparse data; and (3) the number of subscribers that a typical mobile network serves is high, which means that attempting to process and store all the data generated could prove difficult. Indeed, it is one of the bottlenecks of big data to handle the extra network traffic generated as a result of intelligent data analysis in real-time (cf. Manyika et al., 2011).

When collecting location data to be used for finding patterns in human mobility (Figure 1), one of the most important quality factors of the data will be at which frequency the subscribers' locations can be sampled. The population sample frequency that can be achieved will vary greatly, depending on the positioning method and data source used, as well as on the capacity of the network. A high sample frequency could be generated by paging each subscriber's mobile phone more frequently but that would drain some battery life from the phone and incur an extra load on the network, increasing the risk of interfering with its capacity. Naturally, if it is possible to retrieve information about location only once per day, say, it will most likely be very difficult to draw any conclusion about the mobility patterns in the population.

Even if a mobile network processes geographical data about all of its subscribers (which in many countries is equal to almost their entire population), the design of the networks and the massive amount of data available, it is difficult to mine and harvest all of this data. Even if it is currently being generated in every mobile network, this data is not being stored in most networks, so in order to do so, data storage capabilities for the base station controllers would need to be added.

*2.3 Privacy*

To strike a balance between the exploitation of network data and privacy, we must first review the concept of privacy, a multi-faceted concept with differing understandings in multiple disciplines that deals with values, interests, and power (Bennett, 2011). It has nonetheless been defined as an unconditional human right, relating to the amount of control that an individual has over body, territory and, thanks to the advent of bureaucratization and information and communication technology (ICT), over information (Nissenbaum, 2009). This interpretation of privacy is particularly individualistic, arising from issues of personal territory and protection against governmental intrusion, and relating to the traditionally recognized desire to balance the power between government and private individuals (Lyon, 2007, pp. 169-176). Sociologists have also contributed to an understanding of privacy from this perspective, e.g., describing boundary regulation mechanisms as fundamental to personality and agency (Altman, 1975). Privacy, along with its cultural idiosyncrasies, is considered within social sciences as a universal value, essential for describing human relationships.

Privacy can also be defined as a public value, "a social construction that we create as we negotiate our relations with others on a daily basis" (Steeves, 2009, p. 193). Nissenbaum (2009) also views privacy as a social and relational value with contextual complexity. This view describes privacy as a fundamental part of social life and political engagement, allowing separation between public and domestic (private) spheres of life. Such discourses are often built around the notion of trust, which can be broken—thus impacting the relationship—if one of the entities fails to honor the contract.

Matters are further complicated when ICT and pervasive data collection are added to the picture, blurring the contexts by complicating information flows and quickly making laws outdated. From the computer science perspective, models grouped into the concept of privacy-enhancing technologies have emerged that seek to keep information to a required minimum, or to encrypt and anonymize data. These approaches have been inspired by legal frameworks—making ICT comply with existing legislation—and by the principles of data minimization, i.e. any personal data that is not necessary for the functioning of a system or a relationship should not be stored or even collected (Cavoukian, 2012). Other approaches seek to understand how users perceive privacy, and many design methodologies have been applied that encourage close participation between designers and users, like user-centered approaches, such as value-sensitive design, contextual design, user-driven innovation, among others (Bylund, Johnson, Lehmuskallio, Seipel, & Tamminen, 2010) .

For the purpose of the work described herein, there is little use in conforming to any single one of these individual perspectives on privacy in isolation. Rather, we argue that there are good reasons for keeping more than one of these perspectives in mind simultaneously, making certain to cover several of the most salient features of how privacy is understood and interpreted by practitioners as well as users (ibid.).

## 3. Method

The goal is to provide details of the lessons learned in the design of a privacy-sensitive system for assisting modelers and designers of Geographical Information Systems (GIS), or other geospatial visualizations, of human mobility, based on telecommunication network data. The research method is inductive, based upon primary sources from within several multi-year research and development projects that have been recently concluded. These covered socio-technical considerations and privacy-sensitive design processes. The development and innovation focus was on careful mining of the data.

A literature study was conducted deductively, chiefly based on secondary sources. Experiences from designing a system for network data-induced mobility patterns were gained in the context of an industrial and academic project ('Consider8 | SICS', 2010), but the inductive conclusions reported rest on several projects (see ('Prima | SICS', 2005, 'SOLID | SICS', 2002). While the primary sources for the research and development within these projects were largely qualitative, there were also tangible outputs in the form of algorithms, and a patent application (cf. the Results section below).

A case study on transportation planning and analysis was also conducted. Here, information about how people travel in an area was exploited as input to planning new roads, new metro lines, changes of bus routes and train timetables, upgrading of railroad crossings, etc. Traffic analysis models were central in all cases, and the project results aimed for producing valuable input into the validation and calibration of such models. The chief business stakeholder in one of the projects mentioned has already conducted business analyses to that effect.

## 4. System Design

### 4.1 Legal Requirements

The most pressing concern for the system designers is to strike a balance between the privacy of the subscribers and the use of the network data, generated as an artifact of their usage. Therefore, one main activity is to chart the range of privacy issues that this data use may expose. Although as previously stated, privacy is a multi-faceted concept with many interpretations, we anchor our privacy requirements on relevant legal frameworks that are detailed here. In this area, privacy is understood as a violation of the condition of liberty, and therefore relevant only when infringed upon. In other words, privacy as a value is only visible when there is a lack of it. This situation leads to a preoccupation among legal scholars and policy makers to define what constitutes a violation of privacy, which in turn has given rise to many definitions of what is considered personal or private information. Comprehensive regulatory frameworks have emerged that define rights and duties of institutions which collect, store, and process personal data towards the subjects to which the data refers. These have been condensed, e.g., under the umbrella of Fair Information Practices (FIPS) (Fair Information Practices in the Electronic Marketplace, 2000). FIPS are defined as a series of guidelines detailing practices that entities collecting personal data should follow in order to balance the needs and benefits of the entities and the rights of the subjects. They embody five common principles: (1) Notice/Awareness – data subjects should be made aware of the nature of data processing; (2) Choice/Consent – data subjects should be allowed to choose how their own personal data is used; (3) Access/Participation – data subjects should have access to the collected data; (4) Integrity/Security – entities collecting data should ensure accuracy and secure the data from unauthorized access; and (5) Enforcement/Redress – there should be mechanisms in place to hold data collecting entities accountable for failure to address any of the these core principles.

The legal framework that concerns electronic communications data is far from being stable. At present, there are several laws being discussed at national and international levels. For simplicity, we will review here the current state of European Data Protection Directive (Directive 95/46/EC) (European Parliament, 1995) and Working Party ('Data Protection Working Party, Article 29', 2009) opinions, which can influence current and future directives. A unique feature of European Directives regarding data protection is that they, unlike their North American counterparts, treat governmental services and the private sector in the same way (Stratford & Stratford, 1998). It is also more conservative and harmonized. We base our privacy requirements upon European data protection law since the projects that the results rest on were conducted in Europe. The inductive research reported here and the legal requirements derived should be considered valid within the member states of the European Union.

The European Data Protection Directive (European Parliament, 1995) defines personal data as "any information relating to an identified or identifiable natural person" (p. 38). Identified traffic data, i.e. traffic data associated with the identity of a subscriber can be collected for billing and localization purposes. Its use is highly restricted, never leaving the operator's secured premises. Several mechanisms protect this data, both technical (like firewalls and encryption) and legal. Laws regulating network data allow for this data to be used only within restrictive non-disclosure agreements with a limited number of third parties and, before the data is released to these third parties, it must be anonymized, removing all personal identifiers. This directive does not make any explicit mention to geopositioned data. However, the Working Party 29, comprised of representatives from the EU member states' data protection authorities, the European Data Protection Supervisor, and the European Commission (which has significant weight on the interpretation and enforcement of EU data protection laws), has classified positioning data derived from base stations as personal data, thus falling under the directive, which in turn follows the FIPS principles ('Data Protection Working Party, Article 29', 2009). This means that an application seeking to use any kind of network data has to clearly inform the users about the purposes for which the data is collected, requesting, and receiving unambiguous informed consent. Users must also be granted access to the data being collected about them and allowed to make corrections to the data, should they consider it inaccurate. Moreover, according to EUDP, the controllers of geospatial infrastructure—in this particular context, the mobile operators—are responsible for providing adequate security. Finally, methods must be available for the users to hold operators accountable, if any of these conditions fail. Directive 2002/58/EC (European Parliament, 2006) was addressed specifically towards the protection of privacy in the electronic communications sector, and is in line with the opinions stated by the WP29 ('Data Protection Working Party, Article 29', 2009). It states that an electronic communication provider must erase all network data once it is no longer needed for communication or billing purposes. An exception to this can be granted for the provision of value-added services, in which case the data must be anonymized once the users have given informed consent, respecting all terms of the EUDP. Furthermore, it states that only the data necessary for the provision of the value-added service should be saved and processed, and that its access should be restricted to persons acting under the authority of the provider or, if it is the case, the third-party providing the value-added service.

The other exception to this is the Data Retention Directive (European Parliament, 2006) which states that all contextual communication data, which includes network data, must be stored for a minimum period of 6 months, and up to 24 months, for "specific public order purposes, i.e. to safeguard national security (i.e., state security), defence, public security or the prevention, investigation, detection and prosecution of criminal offences or of unauthorised use of the electronic communications systems". (p. 1). Such a withdrawal of protection from the data is only allowed by authorities with "necessary, appropriate, and proportionate measure within a democratic society" (*ibid.*). It is important to clarify that the EU Data Retention Directive does not provide for reimbursement of the providers of electronic communications for the demonstrated additional costs incurred when storing and managing enormous quantities of network data for such long periods of time. Although the EU has recognized the opinion that member states can offer compensation for these additional security measures, they are not forced to do so. This creates pressure for monetizing the data as a way of covering the operator's costs for compliance, and thus can be considered an important driver for capitalization of such data. This might generate interest to systems similar to the one we present here.

In summary, the laws and regulations of the European Union provided a structure for implementing privacy in our system. Although far from being comprehensive, these laws and the FIPS to which they are anchored provide a reasonable framework for processing network data. They clearly indicate that traffic data can be used for value-added services by authorized third parties, provided that all personally identifiable information is *anonymized*. It is from this premise that our methods and algorithms were developed, based on a case study that we detail below.

*4.2 Case Study: Transportation Planning and Analysis*

Transportation systems play a significant role in the growth and development of a society (Bardi, Coyle, & Novack, 2006; Hurst, 1973) with transportation planning and analysis aiming to design an efficient and adequate infrastructure. Hence, transportation planning authorities are interested in analyzing and understanding the human mobility patterns and requirements of the end-users within their geographical areas. However, current methods for acquiring such data (travel surveys, number plate recognition, GPS, etc.), share several limitations:

- Only small, and often biased, population samples are considered for analysis.
- Data analysis is slow, and not capable of being processed in real-time.
- High levels of associated costs.

Such factors ultimately restrict the ability of transportation planners and analyzers to extensively monitor and improve the transportation systems on a regular basis. Additional purchasers include advertisement and marketing agencies, security planning authorities, and eco-system development authorities.

The Origin-Destination (O-D) matrix is one of the most important methods of understanding human mobility patterns in transportation planning and analysis (Hurst, 1973; Rodrigue, 2009). One O-D matrix is composed by several zonal pairs where each pair, represented as a cell in a matrix, has information on the flow—i.e., the number of travelers or amount of freight—originating from one zone to another. This analysis, when consistently accumulated across an area's population, presents realistic transportation and traffic models, such as the Spatial Interactions Model (Hurst, 1973). These help determine the supply and demand between various origins and destinations, while also enabling monitoring of the effects of various policy implementations affecting the transportation systems. While O-D matrices are essential in the development of transportation and traffic models, they can also help environmental researchers, marketing agencies, public safety experts, land-use planners, among others. This method can allow for analyses of larger population samples compared to the other existing techniques. The project solution would partially replace and otherwise complement currently used methods to achieve the required information, such as travel pattern surveys, GPS probes, and traffic cameras.

4.2.1 Privacy Requirements

Central to the idea of privacy is maintaining the complete anonymity of each individual represented in the derived mobility patterns. This anonymity requirement developed from the European data protection regulations that refer to network data. Network data contains location information which, in relation to anonymity, is quite complex to approach from a data-centered perspective; the location of an individual can be considered a formal identifier, especially if we consider real-time applications, while for most cases, location is a quasi-identifier, which means that even after removing identifier data—e.g., full name, government ID number, or telephone number—location can still be used to identify individuals and track them individually. It is worth repeating the definitions of identifiers (Elliot, Hundepool, Nordholt, Tambay, & Wende, 2009):

- Formal identifier: Any variable or set of variables which is structurally unique for every population unit, for example a population registration number.
- Quasi-identifier: Variable values or combinations of variable values within a dataset that are not structurally unique but might be empirically unique and therefore in principle uniquely identify a population unit.

The primary concern is to make it impossible to pinpoint the identity of any subscriber. Our privacy policy follows from these definitions:

- The subscriber number is considered a formal identifier and should be removed.
- The location of an individual is considered a quasi-identifier and should therefore be aggregated.
- All demographic data is considered quasi-identifiers and should thus be aggregated.

There are several ways to concretize our scenarios and serve market needs while considering privacy aspects. To do that, there are several technical solutions that refer to the data ultimately generated by the system. The most radical take on the anonymization problem is to discard the real data all together, and use the statistical properties of the population to generate a new *synthetic population*. This collection of artificial individuals is generated stochastically, with their macroscopic properties adhering to those of the original population (Voas & Williamson, 2000). Since no connection to real individuals can be made at the micro level, there is no risk for compromising the privacy of an individual. The data set itself is also non-sensitive in that it requires no ethical permission to create, maintain, or freely share. The drawbacks include the loss of links to individuals, the possibility that local interaction among the real individuals becomes lost in translation, and that heterogeneous and feedback properties are somewhat unaccounted for.

Table 1. Different datasets that can be constructed from raw network data

| Dataset | Anonymity | Appropriateness for the case study |
|---|---|---|
| Synthetic populations | N/A | No link to individuals and loss of microscopic characteristics of population, therefore not appropriate. |
| Density graph | All personal identifiers are removed and location data is kept. | Link to individuals is weak. Possible to infer frequently traveled routes but not origin/destination pairs. |
| Pseudonymized | Personal identifiers are replaced with a random identifier, consistently used across the dataset. It is prone to inference attacks | It would be appropriate for the case study. |
| Cloaked regions | Locations are blurred (e.g. a location point is transformed into an area). Even if personal identifiers are pseudonymized, it is prone to inference attacks (Khoshgozaran, Shahabi & Shirani-Mehr, 2011). | Dependent on the cloaking level. |

When dealing with real subscriber data, one of the basic attempts to make network data anonymous consists of simply removing a formal identifier from the dataset, replacing it with a unique randomly generated identifier, a *pseudonym*. This technique, known as *pseudo-anonymization*, suffers from a weakness that (Jin, LeFevre, & Patel, 2009) call *temporal linkability*. This means that knowing that a subscriber spends most nights of the week at one address and spends most work hours at another can be enough to identify most people with regular commuting schedules, if aided by auxiliary information, e.g., telephone directory listings.

Replacing pseudonyms at random times would make it harder to reconstruct individual trajectories. Although this would confer some degree of anonymity, it has been shown that it is still possible to attack such a dataset and track particular users across time (Fung, Wang, Chen, & Yu, 2010). *Ad absurdum*, one could use this technique to replace the pseudonym at every point in time, which basically produces a *density graph*, i.e., single non-linkable geographic points in time. Although this would be a privacy-sensitive solution, it is of limited use to the case study we proposed to address. To build meaningful trajectories, such as common routes taken, one needs to track an individual for longer periods of time.

For location-based services designed to be used and accessed in real-time, there are a variety of algorithms that do not preserve trajectories and instead propose to replace single points in time corresponding to positions of individual users with *cloaking regions*, where each region contains a certain number of users (Abul, Bonchi, & Nanni, 2008; Jin et al., 2009).

However, in the case of mobile network data, we are dealing with a different problem than that of a typical location-based service: it is the problem of generating and distributing a geospatial dataset. At this point, for simplicity, we need not worry if the dataset is sensitive as long as it does not leave the operator's premises, where the data resides. The problem then becomes how to properly sanitize this dataset before releasing it to third parties. This approach allows us to construct full individual trajectories from network data which are then anonymized. We discovered techniques that seek to minimize the identification risk after a personal identifier is removed within *privacy-preserving data publishing* which deals precisely with our problem (Fung et al., 2010). These techniques seek to suppress, randomize or aggregate quasi-identifiers, making the dataset safer to use. To achieve anonymity, we start by aggregating the locations of several individuals. This could be done by averaging over geographical areas, similar to most of the previous work we have reviewed. But this would mask characteristics of human mobility, such as routes that are commonly taken, common origins, destinations, and intersection points. Below, we present how this tension was resolved.

4.2.2 Theoretical Underpinnings

To maintain geographical information, we turned to a theory of human mobility developed by Torsten Hägerstrand called *time geography* (1953). Hägerstrand's mission was to represent individual behavior through the inseparable dimensions of space and time in socio-economic systems, and to do so he proposed several constructs.

A time-space *path* represents simply a unique physical path taken by an individual during a certain time. In turn, this path concretizes a prism, a representation of a set of all points that an individual can reach at a certain speed from a certain initial point: a potential-path area. When two individuals occupy the same area at the same time, this shared space can be considered at different levels of scale, including neighborhoods, regional, national, or global. Such time-space aggregates are called *bundles*. Bundling can also occur during travel: an airplane flight or a bus trip is actually an activity bundle—the overlap of the histories of individuals constituted by the convergence of their time-space paths

*Stations* and *Domains* are places where activity occurs, such as physically fixed buildings or territorial units of observation. Examples of these are commercial centers, concert arenas, or bus stations. Stations are flexible and the extent of a station depends on the temporal scale of observation. For example, an individual's home city may be regarded as a station at the life-path scale of observation, but at the daily scale of observation that city is broken up into a number of stations (Holly, 1976) as quoted in (Pred, 1977). Because of its breadth and flexibility, many fields have tried to adopt time geography, either in part or whole cloth, to serve their various purposes. It has primarily been used for obvious applications such as trip planning (Winter & Raubal, 2006) and socioeconomic phenomena (Kwan, 2000). The relationship to location-oriented services is obvious, particularly in the first area. Other uses have come from the fields of sociology and computer simulations (Boman & Holm, 2004).

A theoretical time-geographical database composed of space-time paths is able to answer questions in a wide variety of applications (Boman & Holm, 2004; Miller, 2004; Raubal, Miller, & Bridwell, 2004; Winter & Raubal, 2006). Moreover, a database of bundled paths and stations would not contain any information regarding specific individuals, and would therefore satisfy the requirements of anonymity: the bundles contain demographic data of individuals adjusted for each application, instead of personally identifiable information.

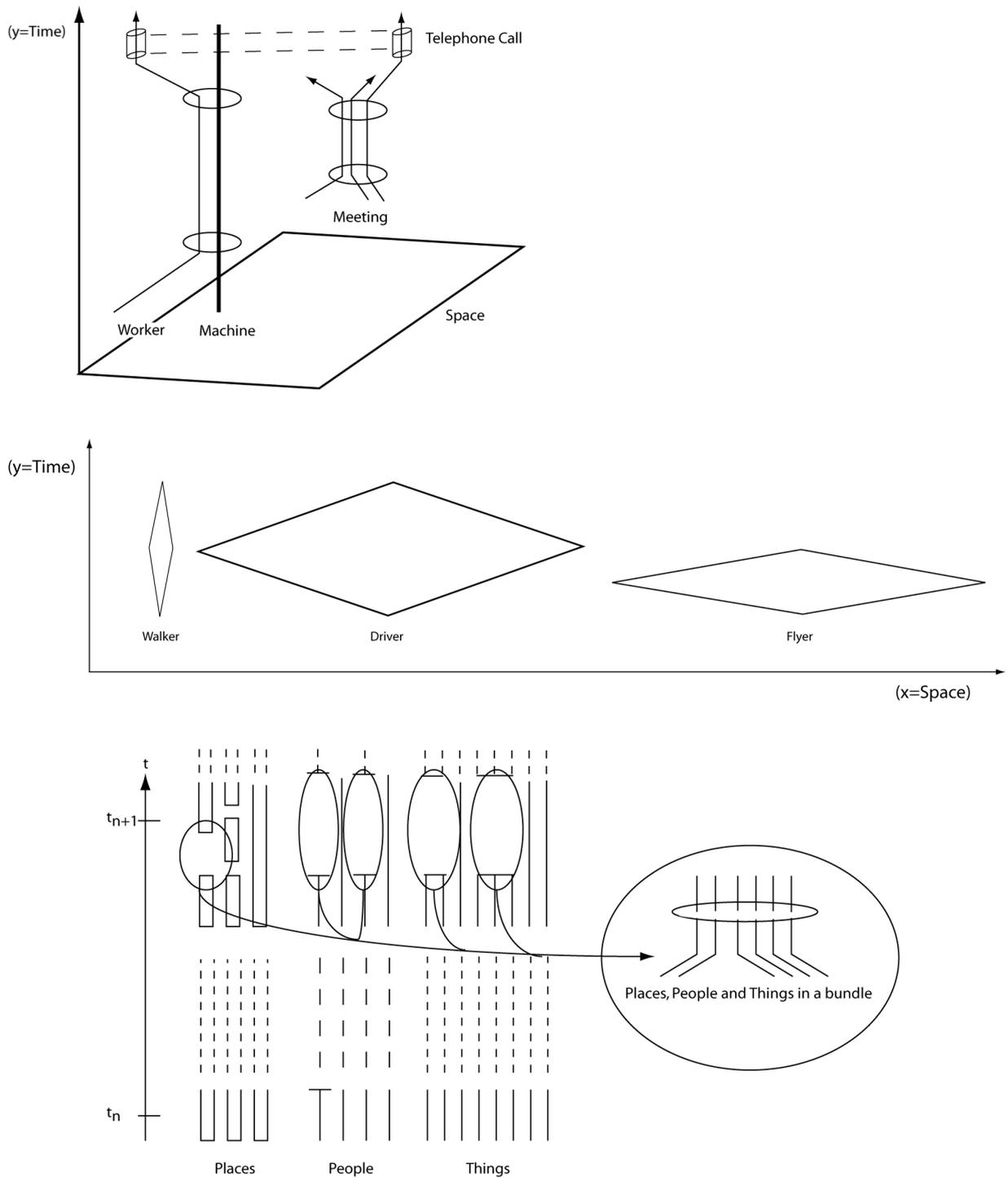

Figure 2. Time-geographical representation of the flow of time (top), prisms representing time-space (middle), and paths being grouped into bundles (bottom)

4.2.3 Proposed Solution

Based on aggregated time-geographical constructs and on an initial study (Svee, Sanches, & Bylund, 2009), the project ('Consider8 | SICS', 2010), a patent application was produced (Bylund, Sanches, & Svee, 2010) that described the technical apparatus of 1) joining two distinct databases within a telecom operator's premises, one pertaining to traffic data, and the other holding customer demographic information; and 2) the general method of aggregation described in terms of time geography, presented below.

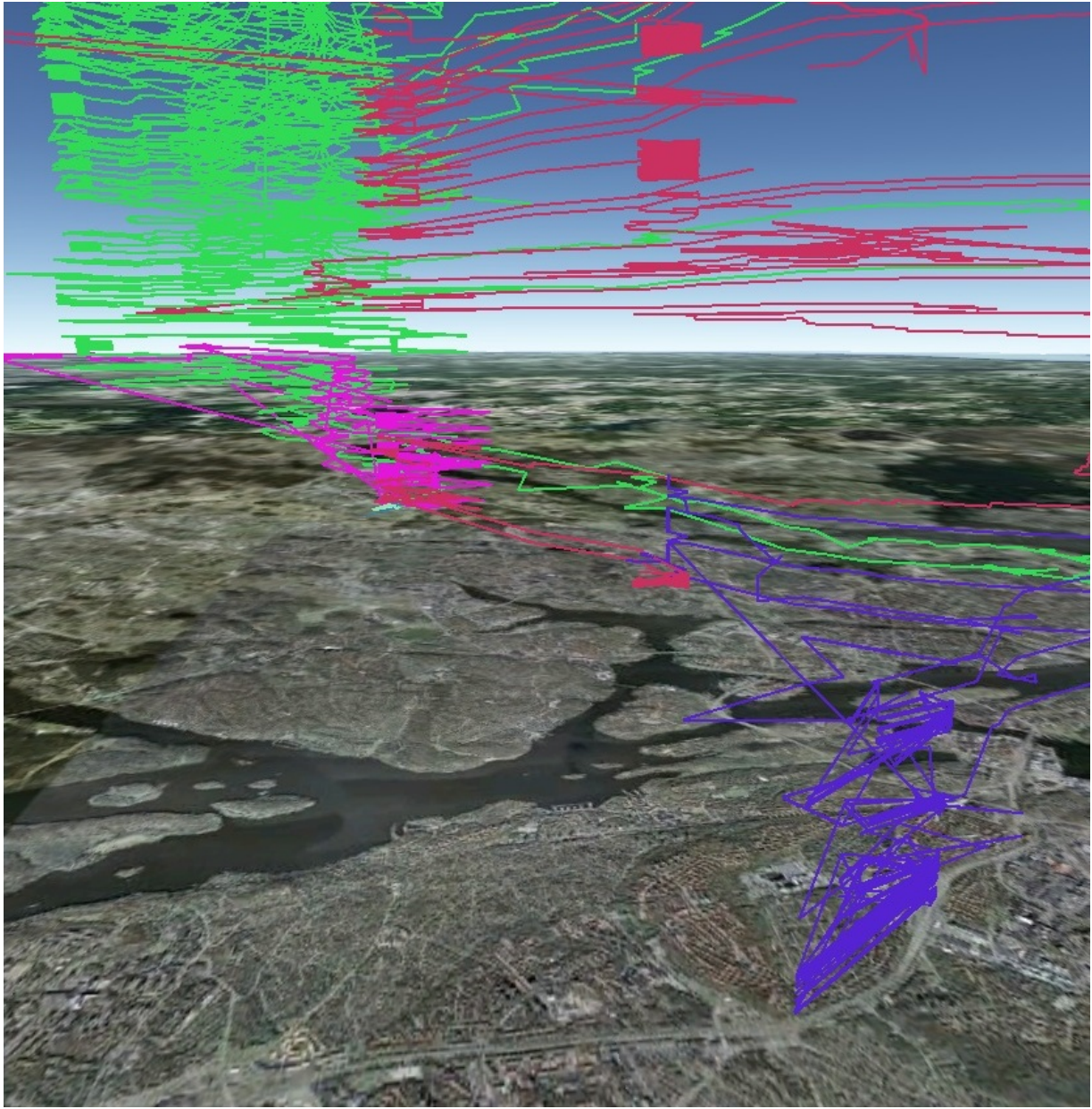

Figure 3. Screenshot of a visualizing tool for our system, depicting a time-geographical representation of four paths, represented by four different colors. The time dimension is represented by altitude

Algorithms and novel methods were devised to transform sparse network data to mobility patterns, which are described in detail elsewhere (Görnerup, 2012; Mellegård, Moritz, & Zahoor, 2011; Nordström, 2012). The proposed system generates statistics about human mobility patterns using information from two GSM subsystems: home location register and customer data management. The information from the home location register is used to determine whether a subscriber is moving or not. In the event they are moving, the information is stored and eventually used to characterize the path the user takes. In the case the user is not moving, the coverage area occupied by the users is registered as a time-geographical *station*. Once a signal characterizing a path or a station is created, it is annotated with demographic information about the individual, held within the customer data subsystem. The annotated signal is then stored in a separate database. Algorithms that convert those sequences of cell-ids to individual space-time paths, with a given duration, origin and destination (Figure 4) are described below.

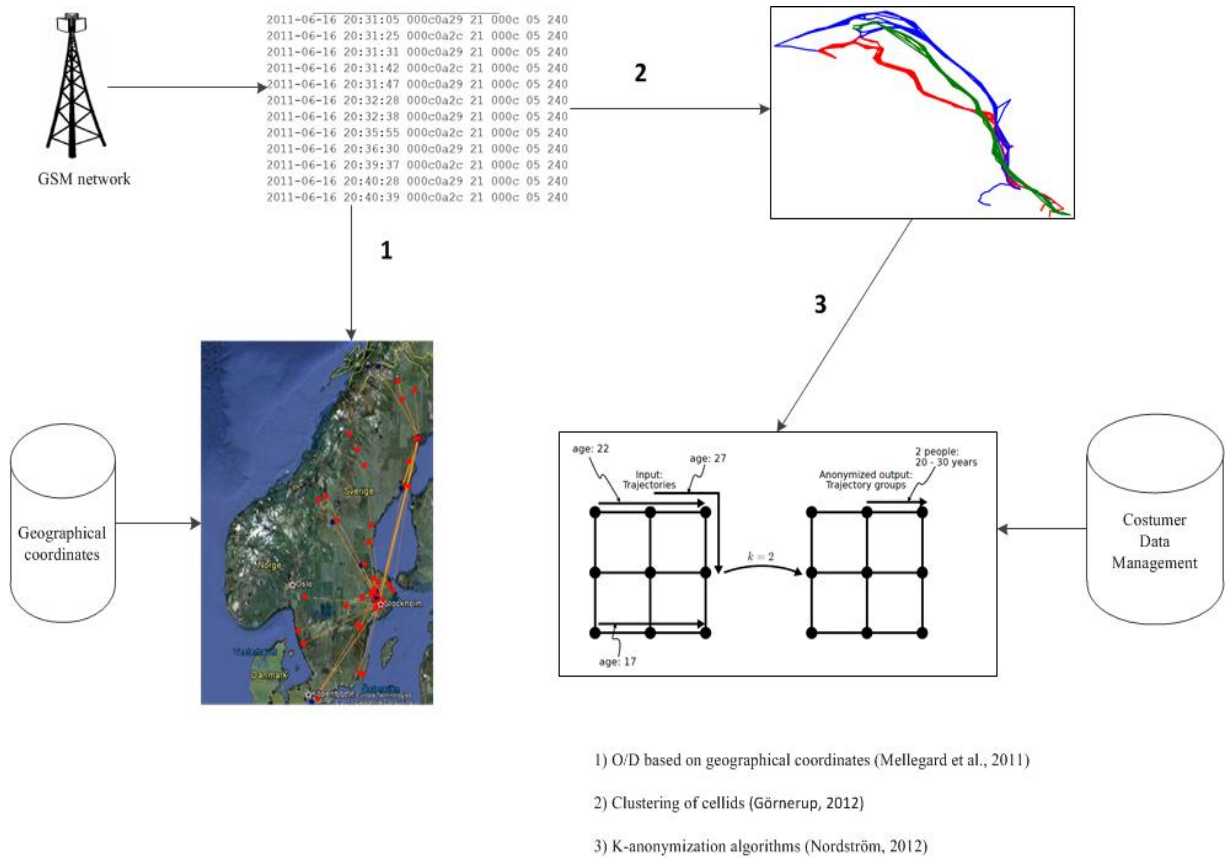

Figure 4. Diagram representing three methods developed within the project: (1) is a simple O/D transformation of raw network data, combined with geographical coordinates; (2) and (3) represent the process of anonymization based on time-geographical constructs, including k-anonymization of network data, and demographic data from the Costumer Data Management of the operator

An initial algorithm was proposed, that uses geo-positioned traffic data to construct O-D matrices, which uses geographical coordinates (Mellegård et al., 2011) obtained through data brokers of cell-id positioning ('Combain | Mobile Positioning with cell id and wifi', 2013, 'OpenStreetMap', 2013) or manually GPS-annotated traffic data, to cluster stations over periods of time. However, there is interest in processing traffic data without the support of geographical coordinates obtained from external entities. For that, researchers tied to our concluded projects have proposed a clustering algorithm (Görnerup, 2012) that processes sequences of cell-ids using a combination of graph clustering (Blondel, Guillaume, Lambiotte, & Lefebvre, 2008) and locality-sensitive hashing. It has been tested on a set of GPS-tagged sequences, and is shown to accurately group sequences. The algorithm scales well with growing number of sequences and is appropriate for usage on large datasets (Görnerup, 2012).

However, as location data is considered a quasi-identifier, additional steps are needed to ensure that the clustered sequences are immune to inference attacks (Sweeney, 2002). To deal with this issue, *k*-anonymization algorithms have been proposed–implementation and tests performed by one researcher tied to our projects are described in detail in (Nordström, 2012)–that divide a sequence in subsequences and discard subsequences represented less than *k* times in the dataset. Generally, *k*-anonymization is a common and efficient method to ensure that each element in the anonymized dataset is undistinguishable from *k-1* elements. This implementation suppresses some subsequences, which means that some information is lost. The algorithms were put to trial in test datasets and achieved linear performance.

At this point the anonymization is not yet complete since demographic data is also considered a quasi-identifier, which means that it may still be possible to identify an individual through inference attacks. For this, we have implemented an interval aggregation algorithm based on the method of *k*-anonymization (Nordström, 2012). In this implementation of *k*-anonymity, no information is suppressed, but rather aggregated and represented as intervals. The combination of these techniques makes the final dataset private, given that the only vulnerability is the size of *k*, in the last steps, and the size of the clusters. These might be dynamic and adjusted to different areas and contexts. Since we have not used access to datasets of real subscribers yet, it is not meaningful to discuss specific values for these variables here.

The final dataset contains no references to individuals. However, while generating the statistics, individually identifiable data is processed. To ensure the integrity of this step, the following measures are taken.

- All data processing is managed by the cellular telephony service operators. This data is already being produced by the network operators; in our system, it is simply being stored and then processed in a novel manner.

- During processing, data used to map sequences onto demographic data (typically a subscriber identifier) is replaced with an identifier that cannot be used to identify the individual (for example, using a one-way hash function). Signals thus annotated are stored separately, thereby reducing the risk of inference attacks on the produced statistics.

The implementation described above also resolves some of the technical issues we identified earlier. By focusing only on important locations—i.e., the most common stations and routes—we indirectly achieve data reduction. The novel sequence clustering algorithm we developed is scalable (Görnerup, 2012) and the final dataset retains a lot of information about the subscribers' mobility patterns, while being significantly smaller than the raw data. We have also shown that, even if the accuracy of the positioning in network data is largely dependent of the physical landscape and the mobile network topology, important locations are still be easily discernible from the cell-id sequences of a large number of subscribers. There is, however, still a possible bias related to accuracy that we have not addressed in our work but will discuss below.

Finally, to initially produce models of human mobility, we had to partially rely on information collected by clients installed on smartphones. Since a mobile device constantly monitors the signal strengths of nearby cells and performs cell reselection for the optimal signal strength, it is possible to use this information to obtain a cell-id with a much higher frequency than through network data. However, the device has no information about the position of each cell so this information has to be retrieved from a database which maps cell-ids to geographical locations. Each mobile operator has a database of the location of each cell, and a number of companies (such as Google, Skyhook, Apple, and Sony Mobile) have also created their own cell-id databases by mapping the current cell-id to a known position, in order to support their location-based services. The reason for using clients in smartphones is the scarcity of network data that makes it difficult to construct a base model. Once the dataset is initialized and a sufficient number of common cell-ids are known, new sparse data can be easily fit to the dataset.

The tests we have performed on synthetic data and data collected from terminal applications recording cell id show great promise of achieving the performance needed to process the volumes of data generated in the networks, while providing rich data and protection from inference attacks. An overall system evaluation to calculate information loss, accuracy, and scalability would be highly relevant, but is meaningful only in a fully assembled system connected to real datasets from telecom operators. In future work, we discuss plans to assemble the system.

## 5. Discussion

The results above are technical components of the system that aim to solve some of the issues we had identified. But as with any socio-technical system, the system is composed of both technical and social elements. Laws and policies specifically targeting network data—as well as other geospatial data relative to individuals—need to be developed to ensure that all implementations of our solution are privacy-sensitive. In this section, we will point out possible solutions for some issues we have identified earlier, based on the (highly Europe-focused) FIPS principles and the opinions of the WP29, as well as identify issues—in particular, some biases—that the implementation of our system raises.

*5.1 Additional Socio-Technical Requirements*

First, it is crucial to document and limit how the data flows as soon as it leaves the operator's premises. We have made the assumption that the data can be freely combined using subscriber identity within the firewalls of the operator, which is what allows for the combination of demographic data from the billing databases with the data from the radio transmitter infrastructure. That communication must be encrypted and access to it should be limited. As soon as it passes through the anonymization steps described above, the data is ready to travel outside of the operator's premises.

We have not yet implemented solutions for the issues of consent and awareness, present in the FIPS. However, given the directions of the legal framework, the operator must allow subscribers to make a choice whether they want to participate or not in data collection, and on which terms. Several options exist, starting by all subscribers and allowing them to opt-out—which might enlarge the sample and be more beneficial for the business model at hand—or encouraging them to opt-in. It should be possible to tailor what kind of demographic information, if any, the subscriber allows to be used in the system. All the data consumers should also be made visible to the subscriber and made aware of what their terms and legal obligations are regarding the data, once transferred to them. Moreover, it should be technically possible for the subscriber to access her own data, before it is aggregated, and prevent it from being shared with third parties, as desired.

Compared to some of the previous modes of obtaining information about traffic congestions—e.g., cameras taking pictures of license numbers to count unique cars passing through an intersection—the method we propose is a more privacy-sensible solution, with lower cost and higher quality. It follows European Union guidelines regarding handling of data and there is no personal data that risk being mishandled or abused. It is important to stress that the contracts binding data consumers and other third-parties accessing the data must be enforceable and transparent, even if the data is anonymized.

*5.2 Biases*

Mining network data poses social risks that are not well captured by the legal frameworks of data protection we have reviewed. One of the biggest is social sorting. Even without demographic data added to the dataset, simply by observing how people move within certain areas, namely where they spend the night and where they spend the day, it is possible to discern class or gender markers, for instance. Different neighborhoods have different demographics and different strata typically have different kinds of jobs. An example that illustrates how things can go wrong from a societal point of view is if we imagine a retailer deciding where to open a shop to sell goods. By looking at the aggregated traffic patterns, the manager can decide to open the shop on the routes where people with higher socio-economic status usually travel through, ignoring the others. This is a phenomenon that has been already identified and discussed by other surveillance specialists (Andrejevic, 2003; Burrows & Gane, 2006). "Postal code sorting" used for marketing purposes has contributed to leaving large, areas, primarily those considered poor, deprived of access to goods and consequently more discriminated against and excluded.

The algorithms developed require data collection on the device for the bootstrapping process, and the only devices that are powerful enough to do so are smartphones. At the moment, Sweden is one of six countries in the world with a population figure over 50 per cent on smartphone penetration (Figure 5). Smartphone users are part of specific demographics in Europe and the U.S: they are better off financially, under 45 years old, and college educated. Therefore, there is a risk for bias, cf. the future case study detailed below. This is a concern since there is a risk that traffic planning disregards elderly, poor, or people of certain ethnicities. Similarly, those who do not use their cellphones during their travels risk misrepresentation in the dataset. To uncover this bias, studies must be done on distributions and demographics of smartphones and usage patterns on the contexts where the system is to operate, paying attention to each country's idiosyncrasies. Yet another source of bias comes from informed consent. If no other participation methods are given to subscribers who chose to opt-out of the network data collection, they risk being excluded from the transportation planning and outcomes as well.

Another important issue concerns the accuracy of maps. Graeff and Loui (2008) point out that technical limitations in GISs can have serious ethical implications. Network data is particularly prone to inaccuracies. Since it is generated by subscriber activity, it typically results in a very low individual sample frequency. The (in)accuracy of geospatial information is a selective inaccuracy, since cities typically have many more cell towers than the country side. That can indicate that a transportation planning system based on network data may be biased and discriminate towards populations located where mobility patterns are difficult to discern. For those, other data collection methods should complement existing ones.

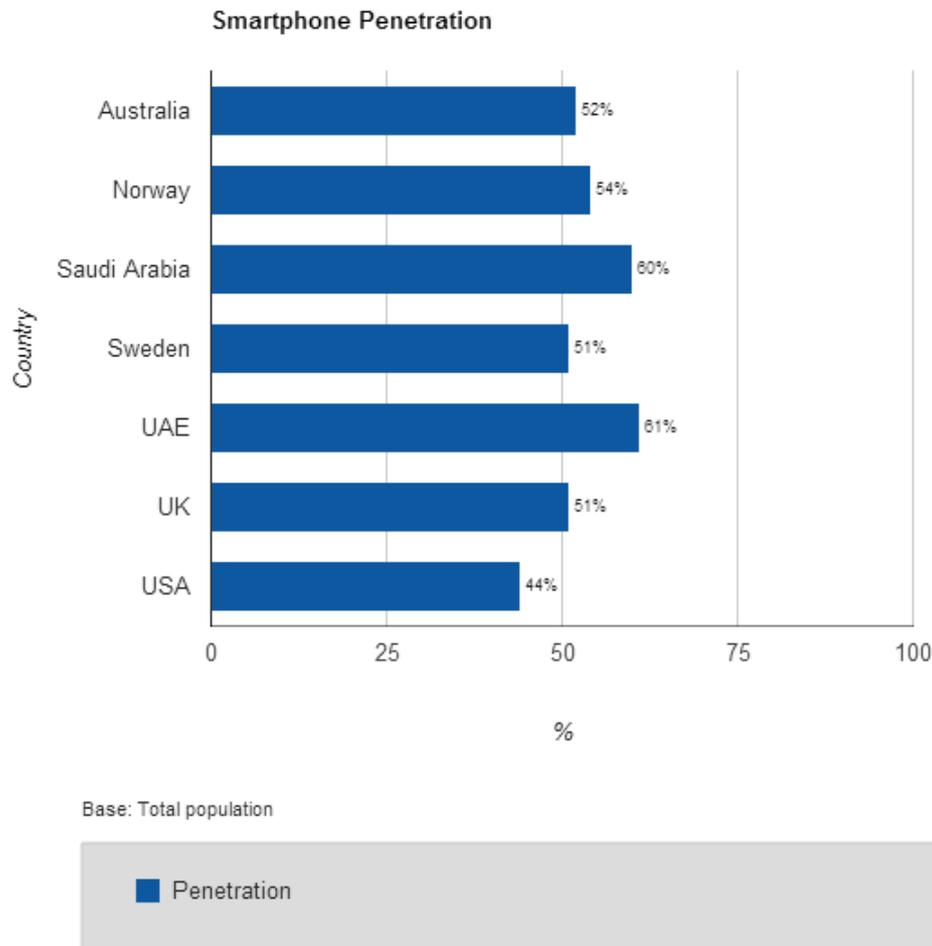

Figure 5. National smartphone penetration rates for the seven top countries in the world; data estimates and graph generated from Google thinkinsights ('Our Mobile Planet', 2013)

*5.3 Future Use Case*

With all the experience gleaned from the case study in mind, we now turn towards proposing a new test case. The idea is to leverage on the lessons learned, avoiding the pitfalls, in a new application. Because this is future research, currently only at proposal stage, the presentation here will be brief, but hopefully detailed enough to indicate why we think it worthy of study from a research perspective.

The problem at hand is the implementation of a syndromic surveillance system in the United Arab Emirates (UAE). This procedure is underway and independent of the future case study proposed here. Such systems use indicators for alerting epidemiologists and other health workers via early signals of the onset of symptoms. We have previously been involved with specifying such a system for the UAE (Boman, Cakici, Guttmann, Al Hosani, & Al Mannaei, 2012). The Health Authorities in Abu Dhabi (HAAD) have an advanced eNotification system (e-IDN, see Health Authority - Abu Dhabi, 2012) for tracking communicable diseases (Underwood, 2009). They are now seeking ways to augment the system with indicators for disease spread, as has been done in the UK (Harcourt et al., 2012), Sweden (Cakici, Hebing, Grünewald, Saretok, & Hulth, 2010), and in many other countries. This is a challenge to do in real-time, because of the input stream of data from a large amount of indicators (Babcock, Babu, Datar, Motwani, & Widom, 2002), but the added value is potentially huge since even a few hours gained on when an epidemic or a catastrophic event is detected can make a substantial difference (Boman, Ghaffar, Liljeros, & Stenhem, 2006).

In the UAE, there is ample opportunity to link access networks, like the ones used for our case study, to the health data available on the population. Such linking has some track record in other countries (see, e.g. Smith et al., 2006). *Etisalat*, one of the 15 largest mobile operators in the world, has the largest market share in the UAE (about seven million subscriptions, compared to six million for competitor *du*, see UAE Interact, 2013), in a country with a population of just over eight million. The emirate of Abu Dhabi has almost complete fiber coverage, and its inhabitants are avid users of commercial and governmental eServices. Hence, the infrastructure allows for bridging the gap between what we in the introduction called 'knowing your population' and the data underlying e-IDN (Koornneef, Robben, Al Seiari, & Al Siksek, 2012).

If one could find respectful ways of employing, with respect to privacy, the micro data in e-IDN, each reported case of sick leave could be instantly linked to geographical and medical data, achieving functional real-time disease surveillance. With indicators for syndromic surveillance in place, the latency in reporting that comes with lab data and the other traditional medical reporting could arguably be overcome. Any future research effort of this kind would also need to investigate the level of trust between the involved parties handling the data; in this case, the health authorities and the country's largest operator. This is important in order to assess the added value of implementing syndromic surveillance. All benefits resulting from augmenting the traditional medical information with non-traditional indicators would also need to be listed, and their weight calculated to counter any challenge to privacy that comes with the implementation.

*5.4 Conclusions*

Aggregated anonymized data is one possible way to address privacy considerations. Our aggregation algorithms, based on time geography, provide reasonable expectations of privacy and if care is taken to follow the Fair Information Practices and general principles of Data Protection, the risk for implementing systems that make use of telecommunication network data is lowered considerably.

Although many have developed technical methods to deal with privacy in electronic records, a thorough analysis of the combined socio-technical consequences is rarely done and many issues are left undocumented. Most of our work has been on how to solve the privacy issues as defined by data protection legal frameworks, but some issues we have pointed out, such as the several biases and the risks of social sorting, are not contemplated by these laws. They deserve specific methods to assess and contain effects that can negatively impact social efficiency and use.

How does one go about dealing pre-emptively with these consequences? When it comes to designing systems that make use of network data, we recommend designers to be transparent about the envisioned uses for the data and allow open discussion to take place where both experts and the general public are encouraged to participate by voicing their opinions and concerns. There have been attempts to create such fora in other emerging technologies (Swierstra & Rip, 2007).

In sharing our experiences and in mapping out a possible practical future use case, we seek to describe an ecosystem of trust, where open dialogue is privileged over proprietary business practices, conditions for the usage of data negotiated, and further requirements easily uncovered. These conditions lower the risks associated with implementing systems that make use of human mobility tracking. They can also contribute to better public use of associated services, ultimately yielding higher public acceptance.

**Acknowledgements**

The authors would like to thank all the researchers involved in the production of the analyses and technical contributions described in this paper. Particularly, we would like to thank Dr. Olof Görnerup, Martin Nordström, Simon Moritz, Richard Carlsson, Erik Mellegård and Marika Stålnacke. We would also like to thank Dr Cother Hajat for important technical assistance. BH and MB would like to thank the director of EBTIC, Nader Azarmi, for the opportunity to investigate the feasibility of the future use case. The work described here was partly supported by the Portuguese funding institution FCT - Fundação para a Ciência e a Tecnologia, with the grant SFRH / BD / 60803 / 2009.